\documentclass[twocolumn]{aastex63}

\usepackage{mathtools}
\usepackage{bm}
\usepackage{esvect}
\usepackage{amsmath}

\begin{document}

\title{Searching for Diamagnetic Blob Accretion in the 74-day K2 Observation of V2400 Ophiuchi}

\shorttitle{K2 observations of V2400 Oph}
\shortauthors{Langford et al.}

\author[0000-0001-5312-649X]{Andrew Langford}
\affiliation{Department of Physics, University of Notre Dame, Notre Dame, IN, 46556, USA}

\author[0000-0001-7746-5795]{Colin Littlefield}
\affiliation{Department of Physics, University of Notre Dame, Notre Dame, IN, 46556, USA}
\affiliation{Department of Astronomy, University of Washington, Seattle, WA, 98195, USA}

\author[0000-0003-4069-2817]{Peter Garnavich}
\affiliation{Department of Physics, University of Notre Dame, Notre Dame, IN, 46556, USA}

\author[0000-0001-6894-6044]{Mark R. Kennedy}
\affiliation{Jodrell Bank Centre for Astrophysics, School of Physics and Astronomy, The University of Manchester, M13 9PL, UK}

\author[0000-0001-5387-7189]{Simone Scaringi}
\affiliation{Centre for Extragalactic Astronomy, Department of Physics, Durham University, South Road, Durham, DH1 3LE, UK}

\author[0000-0003-4373-7777]{Paula Szkody}
\affiliation{Department of Astronomy, University of Washington, Seattle, WA, 98195, USA}

\correspondingauthor{Andrew Langford}
\email{alangfor@nd.edu}

\submitjournal{The Astronomical Journal}

\received{2021 August 18}
\revised{2021 October 11}
\accepted{2021 October 12}
\published{2021 December 10}

\begin{abstract}

Since its discovery in 1995, V2400 Ophiuchi (V2400 Oph) has stood apart from most known intermediate polar cataclysmic variables due to its proposed magnetic field strength (9-27 MG) and disk-less accretion. To date, the exact accretion mechanism of the system is still unknown, and standard accretion models fail to accurately predict the peculiar behavior of its lightcurve. We present the K2 Campaign~11 light curve of V2400 Oph recording 74.19 days of photometric data cadenced at 1 minute. The light curve is dominated by aperiodic flickering and quasi-periodic oscillations, which make the beat and spin signals inconspicuous on short timescales. Notably, a log-log full power spectrum shows a break frequency at $\sim10^2$ cycles~d$^{-1}$ similar to some disk-fed systems. Through power spectral analysis, the beat and spin periods are measured as $1003.4\pm0.2$ seconds and $ 927.7\pm 0.1$ seconds respectively. A power spectrum of the entire K2 observation demonstrates beat period dominance. However, time-resolved power spectra reveals a strong dependence between observation length and the dominant frequency of the light curve. For short observations (2-12 hrs) the beat, spin, or first beat harmonic can be observed as the dominant periodic signal. Such incoherence and variability indicate a dynamical accretion system more complex than current intermediate polar theories can explain. We propose that a diamagnetic blob accretion model may serve as a plausible explanation for the accretion mechanism. 

\end{abstract}


\section{Introduction}
Cataclysmic variable stars (CVs) offer astronomers the rare chance to observe accretion events over very short time scales. CV systems consist of two closely orbiting stars, a white dwarf primary and a red dwarf companion. As the companion star's radius exceeds its Roche lobe, mass begins to fall onto the white dwarf (WD), producing electromagnetic radiation in excess of the binary system's stellar photospheric emission. Periodicities in the binary system, such as the orbital period (P$_{\Omega}$) or WD spin period (P$_{\omega}$), often create observable periodic signatures in light curves. These brightness variations offer a probe for characterizing accretion processes in the binary system.

CVs that contain WDs with significant ($>1$ MG) magnetic fields fall into two categories: polars, the strongest magnetic WDs, and intermediate polars (IPs), more moderate magnetic WDs. Standard IP theory predicts that if a white dwarf possesses a moderate magnetic field strength ($\sim 1-10$ MG), an accretion disk can form around the white dwarf \citep{Patterson1994}. The light curve of such a system should usually show a signal from the spin period of the rotating WD. The azimuthal symmetry of the disk enables both magnetic poles of the WD to be fed simultaneously via accretion curtains, which are luminous structures formed by plasma flowing along the WD's field lines. The curtains co-rotate with the WD, inducing optical fluctuations as the projected area of the curtain facing the observer changes periodically \citep{HellierBook}.

In the realm of magnetic cataclysmic variables, V2400 Oph\footnote{The Gaia EDR3 \citep{edr3} distance to V2400~Oph is $700^{+11}_{-9}$~pc \citep{BJ21}.} (RX 1712.6-2414) holds its place as a strikingly peculiar object. Since its discovery in the 1990s by \citet{Buckley1995Discovery}, V2400 Oph has been identified as one of the most strongly magnetized \citep[$\sim 9-27$ MG;][]{vaeth} intermediate polars (IPs). An outlier of the IP population, the strength of V2400 Oph's spin pulse in optical photometry is marginal compared to the spin-orbit beat (synodic) period. For this reason, V2400 Oph $(P_{\omega} \sim 927 \ \text{s},\ P_{\Omega} \sim 3.41 \ \text{h})$ is commonly cited as the most convincing example of a persistently diskless\footnote{The terms ``diskless IP'' and ``stream-fed'' IP are used interchangeably in most literature, and we follow this convention throughout this paper.} IP \citep{buckley1997}. In a diskless IP, the accretion stream from the donor star feeds whichever magnetic pole is nearest, and the WD's rotation will cause accretion to alternate between different magnetic poles \citep{FerrarioModel}. Known as `pole flipping', this process is present in V2400 Oph; however, X-ray data implies that only $25\%$ of the accretion stream participates in pole flipping \citep{Hellier2002}. This evidence suggests there is some continual flow to each magnetic pole at all times, similar to a disk-fed accretion system.

\citet{Joshi2019} provide the most recent observational study of V2400~Oph. Their \textit{Suzaku} X-ray observations in 2014 showed the spin period to dominate the power spectrum, indicating the system had changed from a stream-fed to disk-fed dominant accretion mechanism. Thus, they disfavor pure stream-fed accretion in V2400~Oph, suggesting instead disk-overflow as the most probable accretion mechanism.

\citet{Joshi2019} also discuss the possibility of an alternative mode of accretion: the diamagnetic blob  mechanism explored theoretically by \citet{King1993} and \citet{WynnKing1995}. This model predicts that in IPs, the accretion flow might consist of discrete blobs traveling along non-Keplerian orbits in the outer regions of the white dwarf's Roche lobe. The orbiting blobs vary in size and density and are disrupted on a magnetic timescale \citep[$t_{mag}$; Eq.~1 in][]{WynnKing1995} that depends in part on their densities. As the density of the blob increases, the magnetic field is less influential on its trajectory, and thus will survive longer. Blobs are expected to orbit the white dwarf several times before being disrupted. In the diamagnetic blob regime, the viscous forces would not have the opportunity to spread the accretion flow into a disk, so there would be a torus consisting of blobs of varying density. Theoretical predictions for observable tests of diamagnetic blob accretion have yet to be fully developed.

The distinct nature of V2400 Oph's accretion mechanism makes the system essential to understand within the context of other CVs. The long 74-day K2 light curve supplies ample data to constrain the accretion mechanism guided by previously proposed models and new investigative techniques.

\section{\textit{K2} Observations}

Launched in 2009, the \textit{Kepler} space telescope observed more than 150,000 stars during its 4 year mission \citep{K2Website}. After 2013, Kepler's extended mission, \textit{K2}, began, providing $ \sim 80$ day observations in one patch of sky along the ecliptic \citep{KeplerPaper}. The photometric data collected, which covers a bandpass of 430-890 nm, is primarily used for exoplanet detection via the transit method. However, the K2 mission supplied an opportunity to study variable stars with uninterrupted, high-cadence photometry. \textit{Kepler} observed only a small number of weakly magnetized CVs and IPs during its lifetime, including FO~Aqr \citep{Kennedy2016}, RZ~Leo \citep{Szkody_2017}, MV~Lyr \citep{Scaringi_2017}, and 1RXS~J180431.1-273932 \citep{Sanne_2021}. 

\textit{Kepler} observed V2400~Oph during campaigns 11-1 and 11-2, occurring from 2016 September 24 through 2016 December 8 ($T_0 \approx 2823.33559 \ \text{BKJD}$, where BKJD is the Barycentric Kepler Julian Date, and is defined as BKJD=BJD-2454833.0). The light curve spans 74.19 days cadenced at $\sim 1 $ min. The light curve was extracted from the target pixel file using {\tt lightkurve} \citep{lightkurve} with a hard quality mask.

\section{Analysis}

\subsection{Periodogram} \label{Sec: Periodograms}

\begin{figure*}[t]
    \centering
    \includegraphics[width = \linewidth]{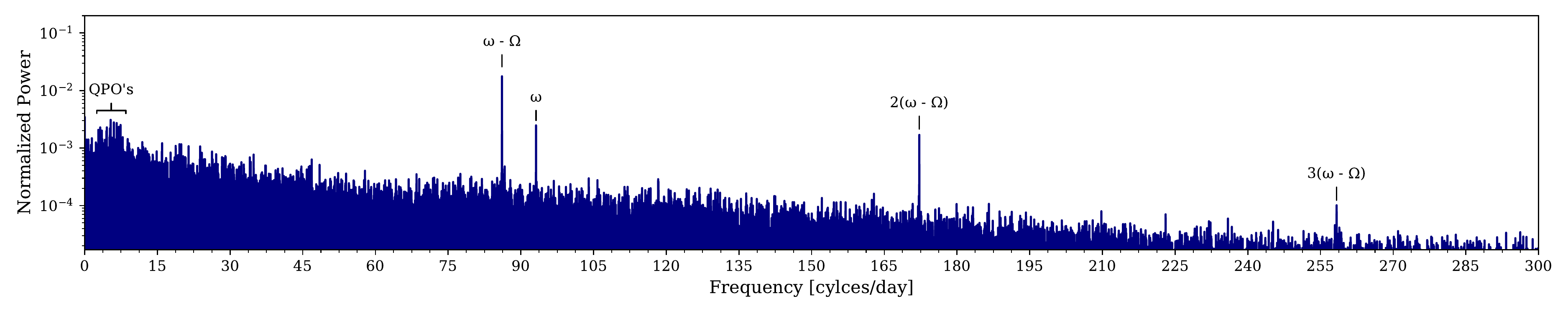}
    \caption{The Lomb-Scargle periodogram of the entire K2 V2400 Oph data set. The notable frequencies, $(\omega - \Omega)$, $(\omega)$, and $2(\omega - \Omega)$, $3(\omega - \Omega)$, are identifiable within the power spectrum. The long duration of the K2 light curve allows for precise measurements of the notable frequencies. Low frequency quasi-periodic oscillations (QPOs) are identified between 2-9 cycles~d$^{-1}$. } 
    \label{fig:ls_periodogram}
\end{figure*}

Unlike many other CVs, the light curve of V2400 Oph shows no coherent periodic fluctuation upon visual inspection of the data. To search for subtle periodic variations, we performed a Lomb-Scargle analysis as implemented in {\tt astropy} \citep{astropy}. This algorithm is useful for astronomical light curves as it does not assume an even sampling rate (\citealt{lombpaper}; \citealt{lombscarglepaper}). To compute the Lomb-Scargle periodogram of the \textit{K2} data, $100,000$ frequency bins from 1/40 to 720 cycles~d$^{-1}$ were assigned a normalized power in accordance with the model, 

\begin{equation}
\begin{aligned}
    & \hspace{70pt} y(t, f, A, C) = \\
    &A + \Sigma^{N}_{n=1} [C_{2n-1} \sin(2\pi nft) +  C_{2n} \cos(2\pi nft)]
\end{aligned}
\end{equation}

where $A$ and $C$ are free parameters to be optimized at each frequency, $f$, and $N = 1$ for single harmonic fitting. 

The normalized power at each frequency is presented in Figure \ref{fig:ls_periodogram}. Although no power is present at the orbital frequency ($\Omega$), the periodogram has well-defined peaks at the spin ($\omega$) and beat ($\omega-\Omega$) frequencies, as well as the first and second harmonics of $\omega-\Omega$. Also, there is enhanced power spread across a cluster of frequencies between $\sim$2-9 cycles~d$^{-1}$, but it is not coherent across the observation. The sharply diminished power below this frequency range suggests that these might be characteristic timescales of variation in V2400~Oph. The lack of power at the orbital frequency ($\Omega=7.024$~cycles~d$^{-1}$) is in agreement with \citet{Buckley1995Discovery} and is explained by the system's low inclination of $\sim 10^{\circ}$ \citep{Hellier2002}.

The spin and beat periods measured from the power spectrum (Table~\ref{table:freqtable}) show no significant change from  \citet{Buckley1995Discovery}; the uncertainties for these periods ($\sim 10^{-1}$ s) were calculated using the technique presented in Sec. 7.4.1 of \citet{vanderPlas}. For comparison, \citet{patterson20} studied five IPs and showed that their spin periods varied by $<0.1$~s over the span of decades, so it is possible that any changes in the spin period of V2400~Oph would go unrecognized in the available data.

\begin{deluxetable}{cc}[]
    \tablecaption{Measured periods from the $ \sim 74-$day K2 V2400 Oph light curve\label{table:freqtable}}
    \tablehead{
    \colhead{$P_{\omega - \Omega}$ [s]} & \colhead{$P_{\omega}$ [s]}
    }
    \startdata
    $1003.4 \pm 0.2 $ & $ 927.7 \pm  0.1  $ 
    \enddata
\end{deluxetable}

\subsection{Trailed Power Spectra}
\begin{figure*}[h]
    \centering
    \includegraphics[width = \linewidth]{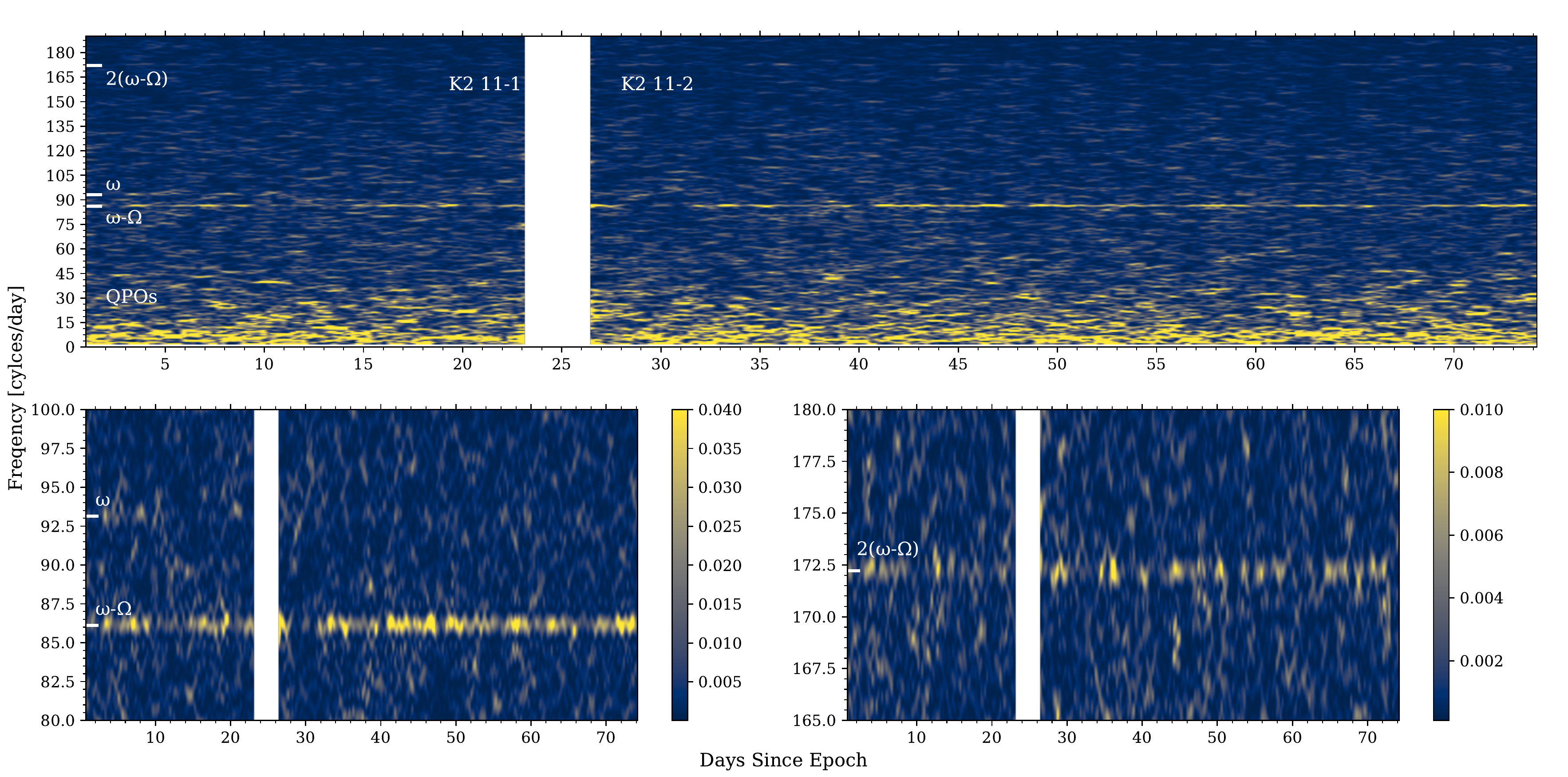}
    \caption{Trailed Lomb-Scargle periodogram of the K2 light curve of V2400 Oph, binned at 1.25 days with 0.25 day step increments. The dominant signals are located within the QPO range, and the amplitudes of the $(\omega - \Omega)$, $(\omega)$, and $2(\omega - \Omega)$ signals all vary sporadically throughout the observation $(T_0 \approx 2823.33559 \ \text{BKJD})$}
    \label{fig:trailed_power}
\end{figure*}

A trailed power spectrum was calculated to study the evolution of the light curve. Figure \ref{fig:trailed_power} shows a trailed power spectrum with a window size of 1.25-days, moved in 0.25-day increments. The frequencies span 10,000 bins from 1.6 to 190 cycles~d$^{-1}$. The consolidated power around the beat frequency at 86.1 cycles~d$^{-1}$ allows the signal to appear dominant and coherent over the lengthy observations. However, the shorter time scale dominant optical fluctuations are occurring between 2-9 cycles~d$^{-1}$ but are unstable, changing frequency on a timescale of days. We identify these as low-coherence quasi-periodic oscillations (QPOs), and in the full periodogram (Figure \ref{fig:ls_periodogram}), their power is spread across a range of frequencies.

Each panel in Figure~\ref{fig:trailed_power} shows the amplitude of the $(\omega - \Omega)$, $(\omega)$, and $2(\omega - \Omega)$ signals to be sporadic throughout the observation. While some variation of signal amplitude is to be expected in a magnetic CV, the level of discontinuity from each signal throughout the observation is a surprising find. Further, there do not appear to be any correlated shifts in power between the $(\omega- \Omega)$, $\omega$, and $2(\omega- \Omega)$ signals.

\subsection{Light Curve and Signal Evolution} \label{Sec: signals}

\begin{figure*}[b]
    \centering
    \includegraphics[width = \linewidth]{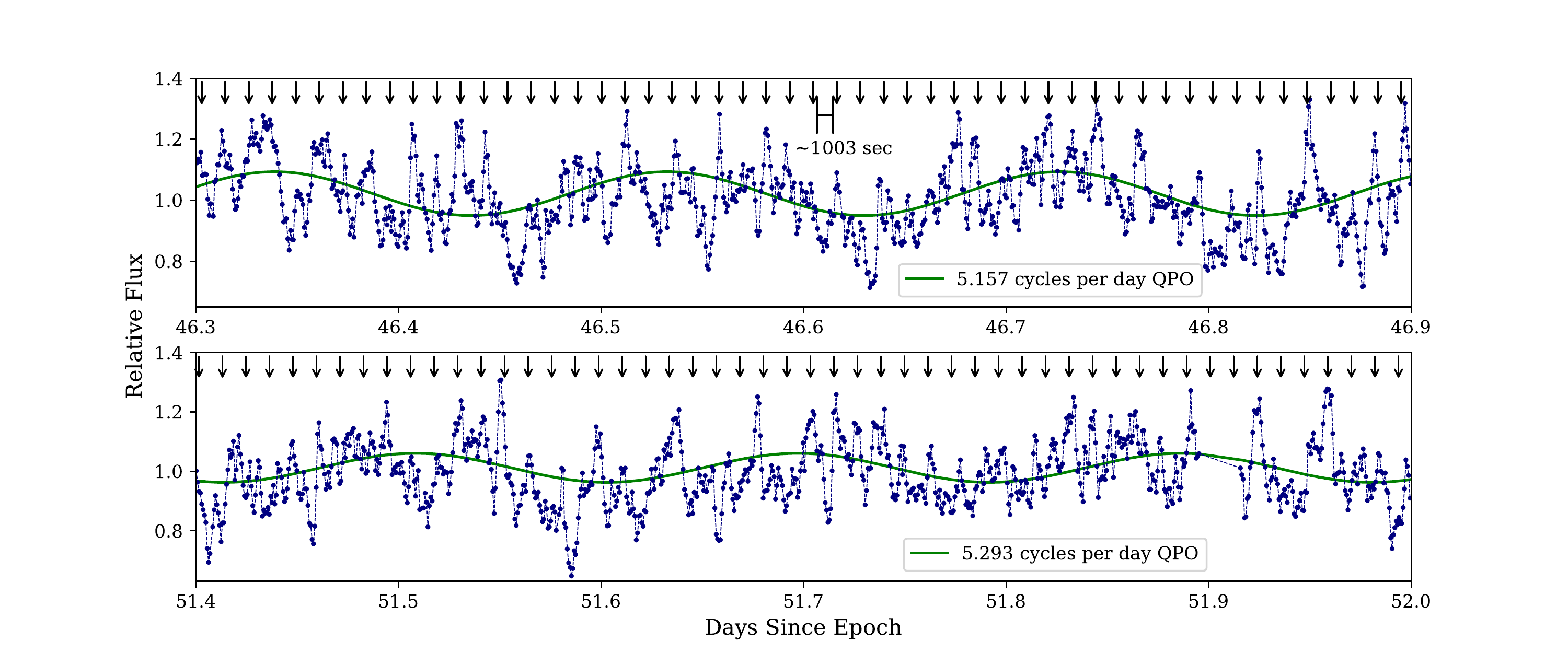}
    \caption{The top and bottom panels show $\sim 14$ hour segments of the K2 V2400 Oph light curve when the beat signal is at $\sim 100\%$ and $\sim 50 \%$ its maximum amplitude respectively. In both segments, a low frequency oscillation at $\sim 5$ cycles d$^{-1}$ is clearly identifiable and accounts for the large flux variations. The K2 V2400 Oph light curve demonstrates more aperiodic fluctuations and chaotic nature than normally observed in CVs. $(T_0 \approx 2823.33559 \ \text{BKJD})$}
    \label{fig:lightcurve_segments}
\end{figure*}

Figure~\ref{fig:lightcurve_segments} shows two 14-hour segments of the light curve at times when the beat signal was at $100\%$ and $50\%$ of its maximum observed amplitude. In both segments, the $\sim 5$ cycles~d$^{-1}$ QPO has a higher amplitude than any periodic signal. The arrows in Figure~\ref{fig:lightcurve_segments} indicate the predicted maxima of the beat signal. Even at the strongest beat signal recorded, the true maxima of the beat signal are difficult to discern. Instead, aperiodic flickering and a low, strong QPO seemingly dominate the light curve.  

\begin{figure*}[t]
    \centering
    \includegraphics[width = \linewidth]{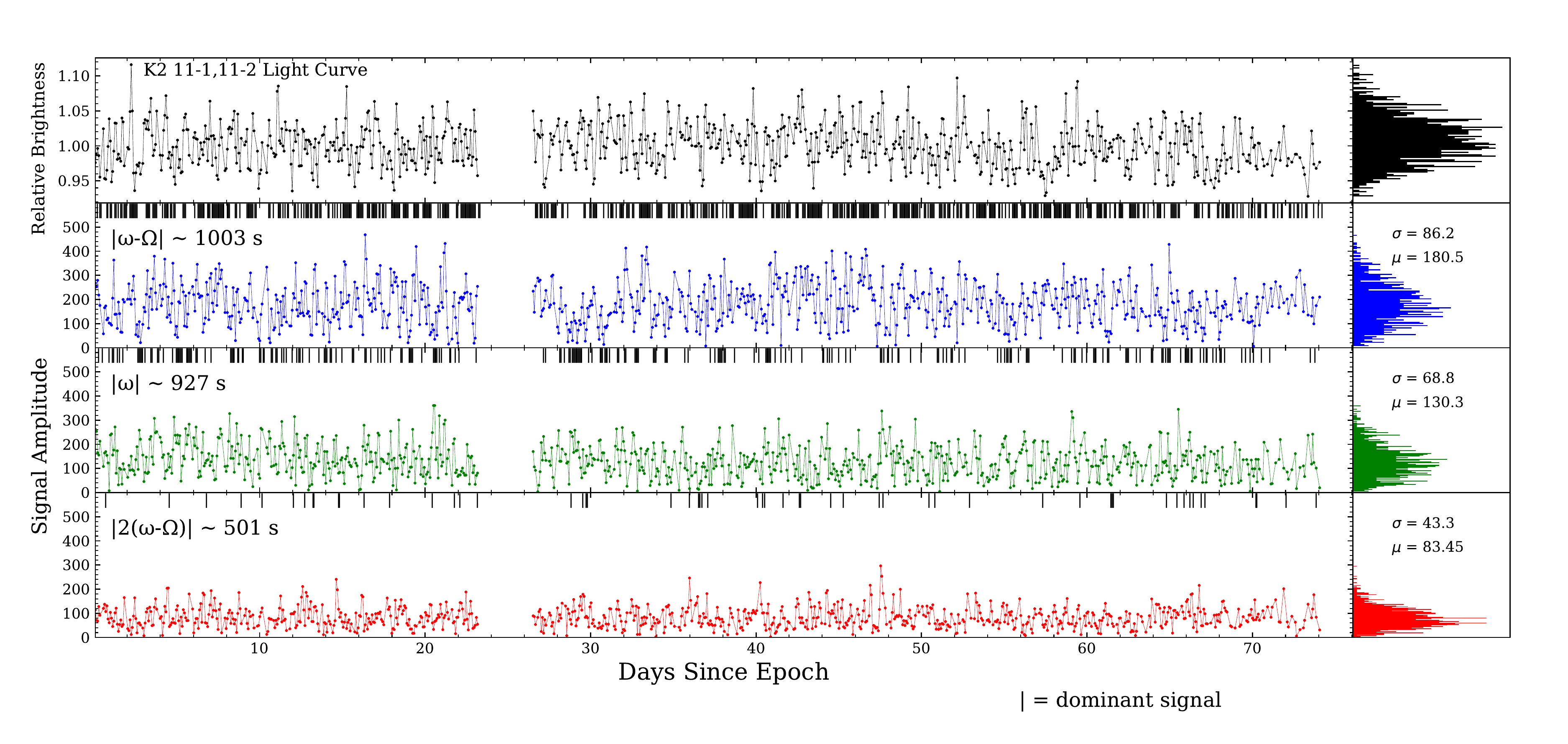}
    \caption{The modulation of notable signals with reference to the relative brightness throughout the K2 V2400 Oph observation. The dominant signal is denoted for each $\sim 5$ hour observation window. Strong variability in the beat signal amplitude allows for both spin and beat harmonic signals to appear dominant for periods of observation. There is no distinct correlation between any of the signals or relative brightness throughout the observation. $(T_0 \approx 2823.33559 \ \text{BKJD})$}
    \label{fig:signal_modulation}
\end{figure*}

Previous studies of V2400 Oph, such as \citet{Buckley1995Discovery} and \citet{Joshi2019}, have observed the dominant frequency in the power spectrum to change between observations. To place these findings in context, Figure \ref{fig:signal_modulation} shows the amplitudes of the $(\omega- \Omega)$, $\omega$, and $2(\omega- \Omega)$ signals over the K2 observation. The bin size (5~h) approximates an average observation time of previous literature on V2400 Oph and a normal observation length for ground telescopes. For each observation window, the signal with the largest amplitude is designated by a vertical tick mark above its average time. While the averaged fractional light curve shows variability, the relative change predominantly stays within 5\% of the average brightness and there are no distinct low or high states throughout the observation --  ruling out dramatic changes in the accretion rate. A comparison of these three signal amplitudes shows that the dominant signal in the power spectrum varies erratically throughout the observation. Of the 851 5-hour bins, 64.9\% were $(\omega- \Omega)$ dominant, 28.5\% were $\omega$ dominant, and 6.6\% were $2(\omega- \Omega)$ dominant. If the behavior during the K2 observation is typical of V2400~Oph across its observational history, this analysis underscores the futility of trying to interpret the system's power spectrum in terms of a dominant frequency -- as is often done for IPs. 

As the bin size increases, the beat frequency is more routinely observed as the highest-amplitude frequency. For a bin size of 1.75~d, the signal dominance changes to 95.4\%, 3.2\%, and 1.4\% for $(\omega- \Omega)$, $\omega$, and $2(\omega- \Omega)$ respectively. It is critical to keep in mind that the low-frequency QPOs are always observed with higher amplitudes than any of these three frequencies for observations less than 5 days.

\begin{figure}[b]
    \centering
    \includegraphics[width = 0.8\linewidth]{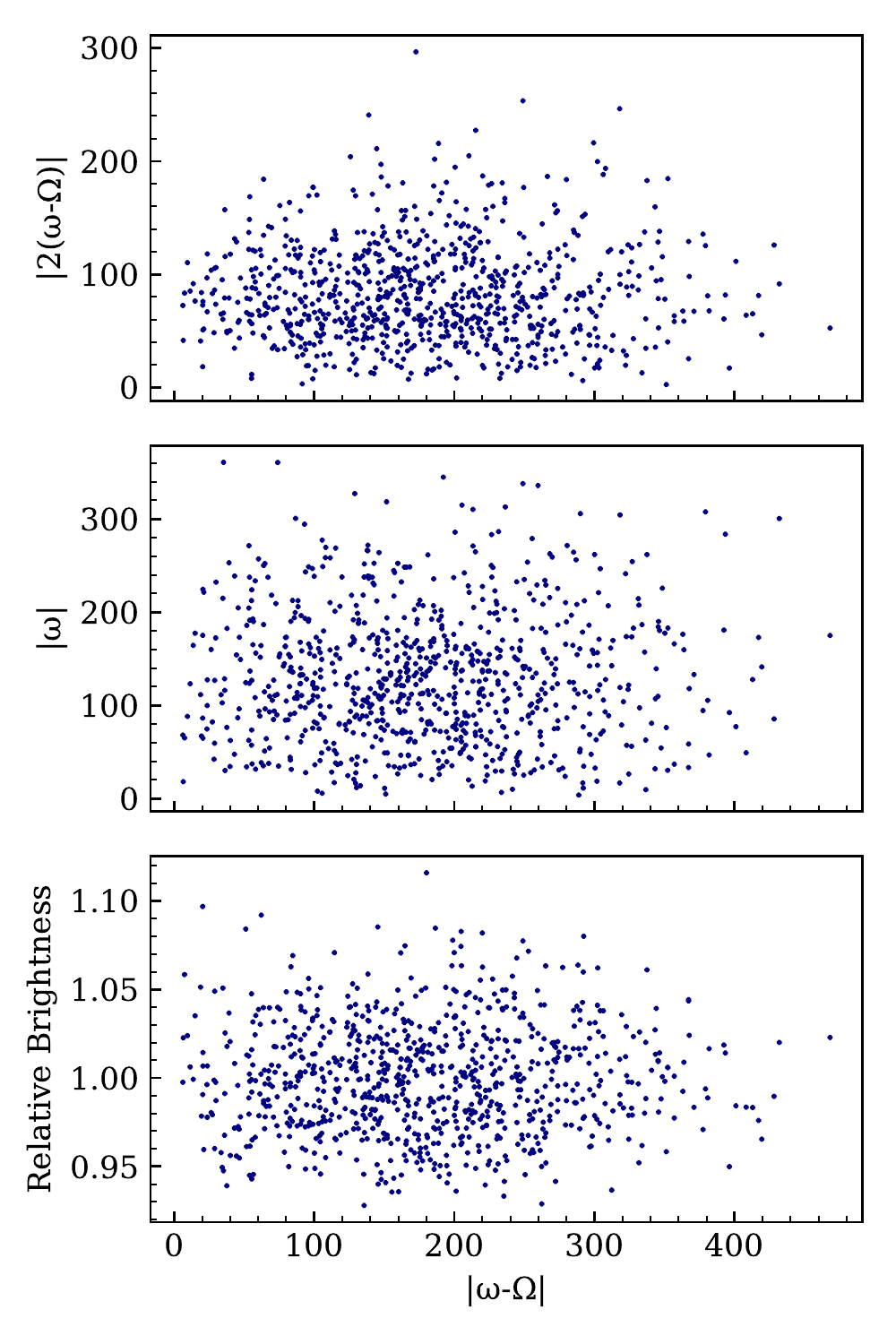}
    \caption{Scatter plots of the K2 data demonstrating the uncorrelated beat signal amplitude with other signals in the system. Each point corresponds with a $\sim 5$ hour observation window.}
    \label{fig:beat_signal_correlate}
\end{figure}
 
Figure \ref{fig:signal_modulation} also demonstrates the variability of each signal's amplitude. Each signal possesses an approximately log normal amplitude distribution. The $(\omega)$ and $2(\omega - \Omega)$ signals show less variance than $(\omega - \Omega)$. The large amplitude swings of $(\omega - \Omega)$ on short time scales may better explain the changes in signal dominance rather than $(\omega)$ and $2(\omega - \Omega)$ becoming significantly stronger for periods of time. Thus over longer observation windows, the $(\omega - \Omega)$ amplitude variability is dampened and there is less chance to observe  $(\omega)$ and $2(\omega - \Omega)$ as the dominant signal. 

Figure~\ref{fig:beat_signal_correlate} shows no correlation between the amplitudes of $\omega$, $\omega-\Omega$, and $2(\omega-\Omega$). It is somewhat surprising to see no correlation between the $(\omega - \Omega)$ and $2(\omega - \Omega)$ signals given their harmonic relation. This may indicate the two signals have separate physical origins. Further, we investigated possibility of and found no phase difference which might correlate the beat signal and the other signals.

\section{Discussion}

\subsection{The Effect of a Face-on Inclination} \label{Sec: Geometry}

Before discussing related observations and theoretical accretion models, it is worth considering the observational implications of the low inclination of V2400 Oph for detecting periodic signals. In a theoretical disk-fed IP system at $\sim 0^{\circ}$ inclination, the observer views the entirety of the disk and both accretion curtains constantly. 

Due to the symmetries of the system as viewed from above, an observer likely would not see the expected optical fluctuations at the spin and orbital frequencies. However, in a stream-fed IP, these symmetries are broken because the accretion flow can flip between magnetic poles. Even at $i=0^{\circ}$, this pole switching can lead to the spin or beat frequencies being readily detectable in power spectral analysis. However, it also possible that even at lower inclinations, a signal may be seen at the spin and beat frequencies due to self-occultation of the post shock region (where the X-ray and the cyclotron emission are generated) by the white dwarf itself \citep{Hellier2002}.
In short, brightness modulations from stream-fed accretion may remain observable at low inclinations, while modulations from disk-fed accretion are unlikely to do so. As a result, a low orbital inclination will likely cause an observational bias in favor of detecting stream-fed accretion in power spectral analysis of optical photometry.

\subsection{Break frequency}

\begin{figure}[t]
    \centering
    \includegraphics[width = \linewidth]{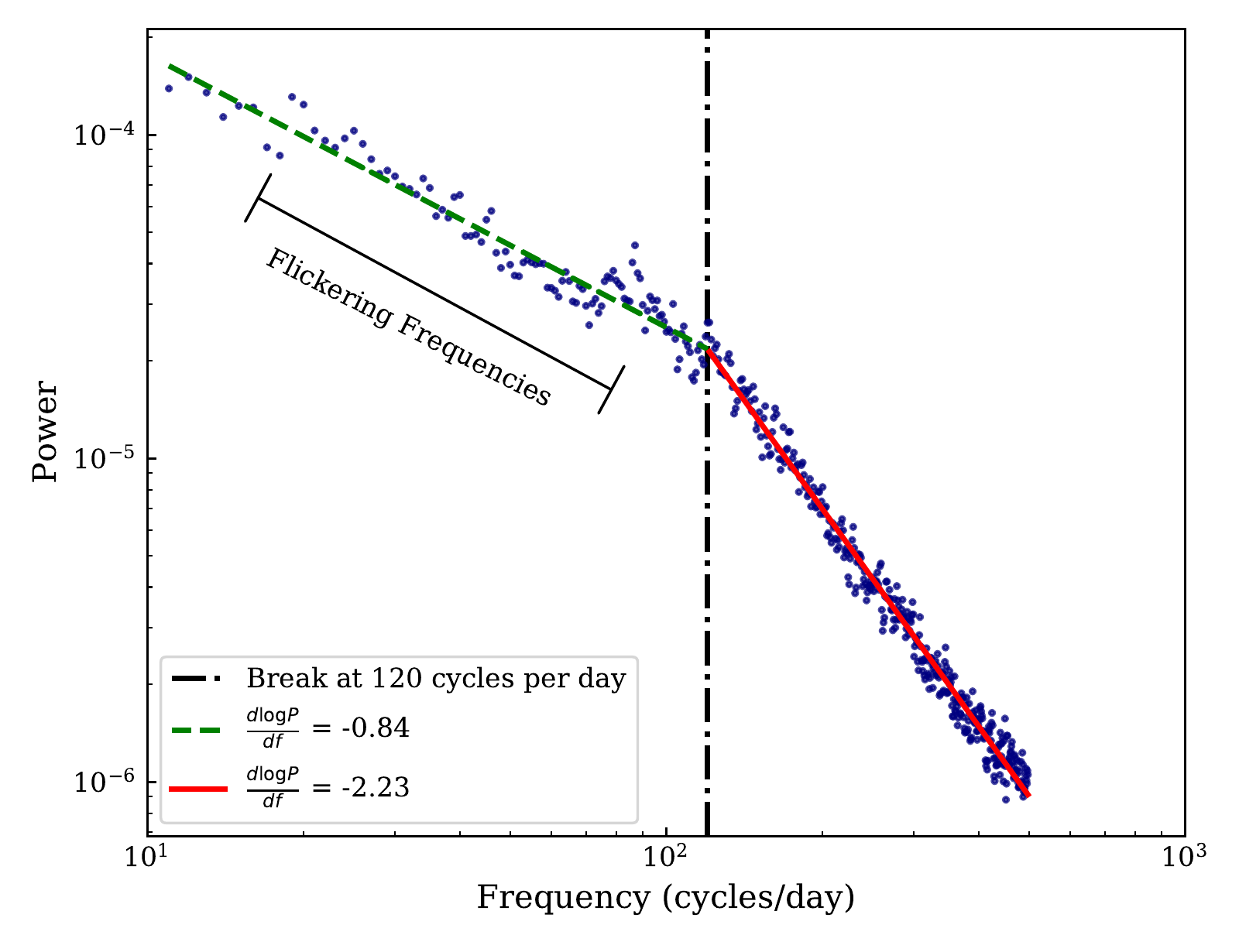}
    \caption{A binned, Lomb-Scargle periodogram of the K2 V2400 Oph data set. The break in slope at $\sim 10^2$ cycles~ d$^{-1}$ is common among disk-fed systems \citep{Scaringi}.}
    \label{fig:loglog_periodogram}
\end{figure}

Recent work has shown properties such as RMS-flux to scale linearly with the energy of accreting systems \citep{Scaringi2016}. The log-log scaled periodogram in Figure \ref{fig:loglog_periodogram} has frequency bins of 1 cycle~d$^{-1}$ and records the average power observed for each frequency bin between 10 to 500 cycles~d$^{-1}$. The result shows a distinct break in slope occurring at 120 cycles~d$^{-1}$. When periodograms are plotted in log-log scaling, disk-fed systems, such as FO Aqr and MV Lyr, are often seen to have breaks around $100$ cycles~d$^{-1}$, and modeling by \cite{Scaringi} shows the aperiodic variability generating the high-frequency break to be associated with a geometrically extended inner accretion flow. Although it has already been discussed how this flow is fundamentally different from a standard geometrically thin disk, it is nonetheless interesting to note that V2400~Oph displays a similar high-frequency break as some disk-fed systems. It is therefore remarkable that V2400~Oph, despite being the prototypical diskless IP, has a power spectrum whose break frequency is comparable to that observed in disk-fed systems.

\subsection{Prior V2400 Oph Observations}

\subsubsection{Initial Optical Observations} \label{Sec: beat_correlation}

V2400 Oph's observed optical emission due to accretion is well in excess to the stellar components. The companion has yet to be detected and WDs are inherently dim due to their lack of surface area. Therefore, the V $\sim$ 5 absolute magnitude (\citet{Shappee, Kochanek}) indicates active accretion processes resulting in optical emission. Rotating accretion curtains, disk-like structures, and cyclotron emission are all expected to contribute to the optical flux in the K2 bandpass. \citet{Hellier2002} deduced that most luminosity sources are circling around the WD.

\cite{Buckley1995Discovery} present the only other in depth analysis of optical photometry for V2400~Oph. Their high-speed photometric observations span from 1991-1994 and average $\sim 2$ hours in length. Initial observations in August and September 1991 recorded an obvious, highly coherent beat pulse. The rest of their optical photometry is similar to the \textit{K2} data in that the periodic variation is overwhelmed by low-frequency, aperiodic variations (see our Fig. \ref{fig:lightcurve_segments}). 

In their power spectral analysis, \cite{Buckley1995Discovery} often found `red noise' characterized by increasing power towards lower frequencies. Our full periodogram in Figure \ref{fig:ls_periodogram} is in good agreement with this observation, and Figure \ref{fig:loglog_periodogram} identifies a distinct break associated with a frequency cut-off for the `red noise'.

The intermittent nature of the beat signal in the K2 data is also in agreement with the \citet{Buckley1995Discovery} observations. \citet{Buckley1995Discovery} found that the amplitude of the beat signal is erratic, ranging from strong to completely absent. As previously mentioned in Section \ref{Sec: signals}, we observe this throughout the \textit{K2} observations, where the beat amplitude demonstrates large discontinuities with no predictable pattern of variation. While the K2 spin amplitude is variable, the distribution's spread is small compared to the amplitude swings observed in the beat signal (Figure \ref{fig:signal_modulation}). Therefore, we presume that the physical phenomena linked to the beat signal must be more dynamically unstable than the source of the spin periodicity. 
 
 In their observations between 1991-1994, \cite{Buckley1995Discovery} found the brightness of V2400 Oph to be anti-correlated with amplitude of the beat signal. The beat pulse in their dataset was well-defined at a brightness of $V\sim14.6$, but when V2400~Oph was 0.5-0.7 mag brighter, the increased aperiodic variability made the beat signal significantly more difficult to detect. For comparison, the $V$-band photometry from the All-Sky Automated Survey for Supernovae \citep[ASAS-SN; ][]{Shappee, Kochanek} establishes that during the \textit{K2} observation, V2400~Oph was relatively bright at $V\sim14.2$. Thus, it is possible that the beat pulse becomes apparent only at lower accretion rates.

\subsubsection{Recent X-Ray Observations} \label{x-ray}
\begin{figure}[t]
    \centering
    \includegraphics[width = \linewidth]{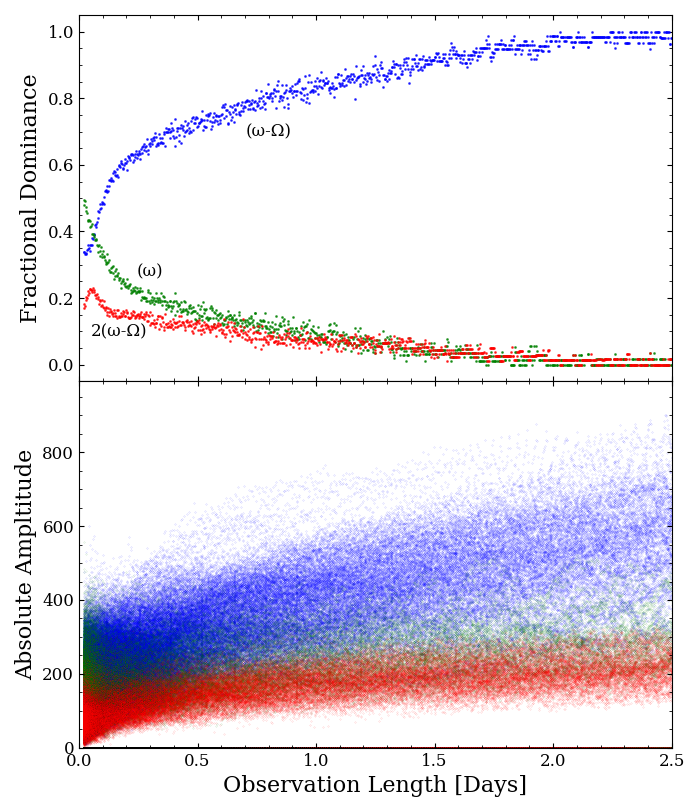}
    \caption{The dominant signal and signal amplitudes as a function of observation window size for the K2 light curve. The probability of observing a given signal with the highest relative amplitude is significantly influenced by the observation length. }
    \label{fig: percent dominant}
\end{figure}

While optical emission in V2400 Oph may come from a range of sources, X-rays are localized to the post-shock region above the WD. This restricts any direct interpretation between X-ray and optical observations.

\citet{Joshi2019} analyzed the beat and spin signals from X-ray observations of V2400 Oph taken by the \textit{XMM-Newton} and \textit{Suzaku} spacecraft. The 2009 and 2014 \textit{Suzaku} data sets provide the most recently published observations of V2400 Oph. Using Lomb-Scargle analysis, \citet{Joshi2019} found the beat and spin signal amplitude to vary between observations and both were nearly absent in 2000. Conversely, the authors noted subsequent observations by \textit{XMM-Newton} (2001) and \textit{Suzaku} (2009, 2014) displayed beat and spin signals. In light of the evolving power spectrum, \citet{Joshi2019} argued that V2400 Oph experienced dominant accretion mode switching between stream- and disk-fed mechanisms. Specifically, \citet{Joshi2019} cited a change in signal dominance from beat to spin between the $1.3$-day-long 2009 and the $0.5$-day-long 2014 \textit{Suzaku} observations. 

The \textit{K2} light curve complicates this interpretation because the relative spin and beat amplitudes are erratic on short timescales. Figure~\ref{fig: percent dominant} demonstrates that the highest-amplitude optical signal depends largely on the duration of an observation. Our analysis in Figure~\ref{fig:signal_modulation} showed that either the spin or beat frequency is commonly observed as the strongest periodic signal for short ($\sim$5~h) observations. However, for observations of longer durations, the beat period becomes the sole dominant period. If the X-ray power spectrum is susceptible to the same effect, it would offer an alternative explanation for the \citet{Joshi2019} conclusion that the mode of accretion changed between X-ray observations.

\subsection{V2400 Oph compared to other IPs}

Of the few IPs observed by \textit{Kepler} or \textit{TESS}, FO~Aqr and TX~Col offer an excellent point of comparison. Both FO Aqr ($i = 65^{\circ} -75^{\circ}$, \cite{Hellier1989}) and TX~Col $(i \lesssim 60^{\circ})$ are well known for having demonstrated variability in both their mass accretion rate and accretion mechanisms, with TX~Col having shown this behavior during its \textit{TESS} observation.

\subsubsection{FO Aqr}

\begin{figure}[t]
    \centering
    \includegraphics[width = \linewidth]{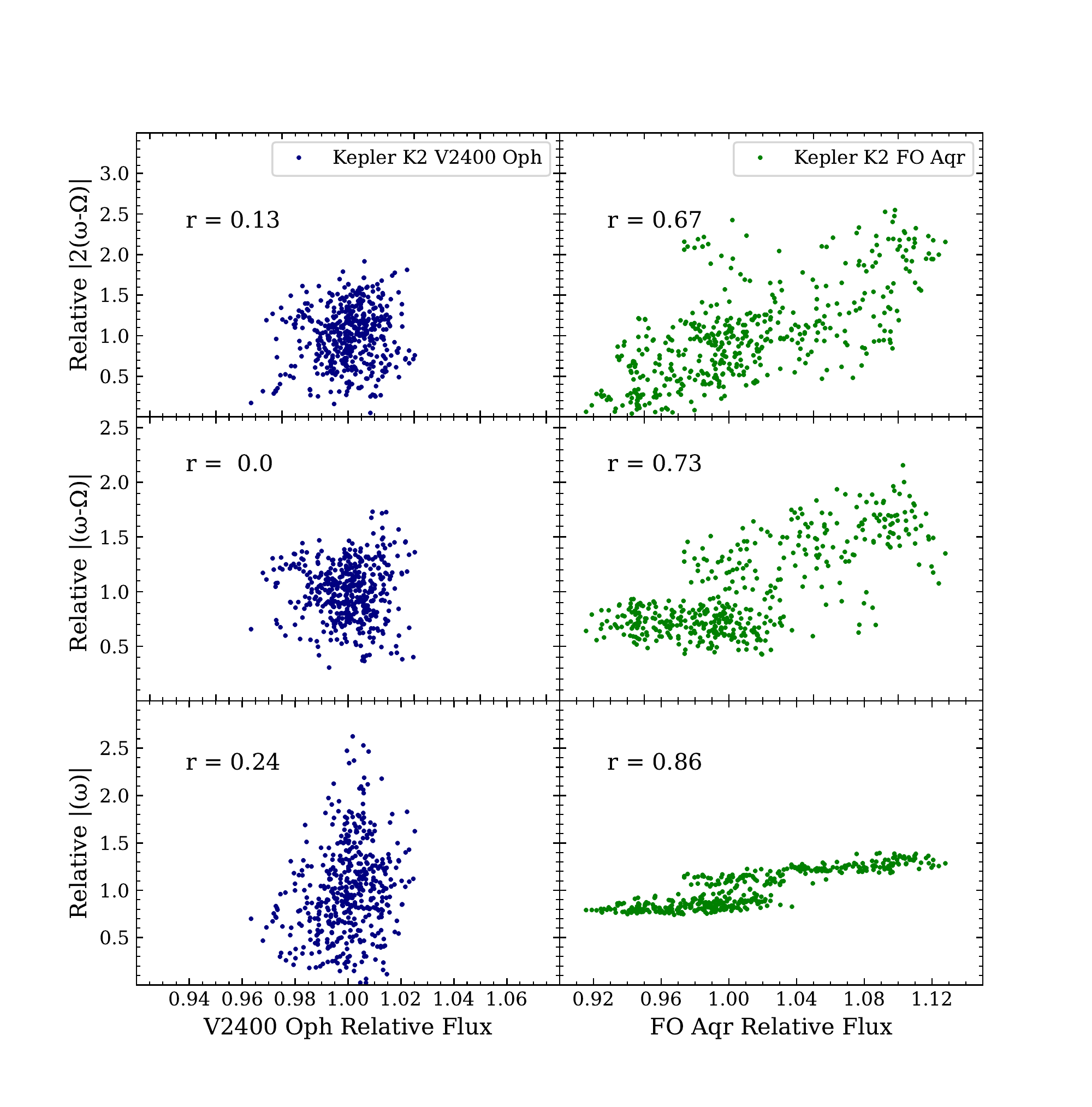}
    \caption{Correlations between relative brightness and signal amplitudes of both V2400 Oph and FO Aqr K2 observations. Each data point corresponds to a 1.5-d observation window. FO Aqr shows some correlation between each signal and the relative brightness of the system. The V2400 Oph signals appear to be uncorrelated with the system's brightness.}
    \label{fig:signal_correlations}
\end{figure}

The 69-day trailed power spectrum of FO~Aqr shown in Figure \ref{fig:fo_trailpower}\footnote{See Appendix} depicts FO Aqr in its high state during which it is a prototypical disk-fed accretion system, and the contrast with V2400 Oph's power spectrum is dramatic. As expected for an eclipsing, disk-fed accretor, FO Aqr's trailed power spectrum is dominated by the spin and orbital frequencies, and there is no evidence of QPOs. In comparison, V2400 Oph is a top IP candidate for disk-less accretion and possesses a trailed power spectra distinct from the well behaved proprieties of FO Aqr. Even if the inclination of V2400 Oph was high enough to produce a large-amplitude spin pulse (Sec. \ref{Sec: Geometry}), the unpredictable amplitude variation is uncharacteristic for well behaved disk-fed systems. 
During its K2 observation period, FO Aqr was found to possess strong correlations between system brightness and signal amplitudes \citep{Kennedy2016}, implying that the relative strengths of the side band frequencies were related to the mass transfer rate. In contrast to V2400 Oph, FO Aqr's beat signal is observed consistently throughout the duration, and \citet{Kennedy2016} associated it with disk-overflow accretion. As shown in Figure \ref{fig:signal_correlations}, the beat amplitude of FO Aqr correlates positively with an increase in the star's brightness. The reliability and coherence of FO Aqr is in stark contrast to the chaotic nature of V2400 Oph. Figure \ref{fig:signal_correlations} demonstrates again the unpredictable behavior of V2400 Oph, which exhibits no correlation between the system's brightness and its three strongest signals. Moreover, unlike V2400 Oph's $|(\omega- \Omega)|$ and $|2(\omega- \Omega)|$ in Figure \ref{fig:beat_signal_correlate}, FO Aqr demonstrates some correlation between its $|(\omega- \Omega)|$ and $|2(\omega- \Omega)|$ signals suggesting that the two signals are produced through the same or related mechanisms. 

\begin{figure}[t]
    \centering
    \includegraphics[width = \linewidth]{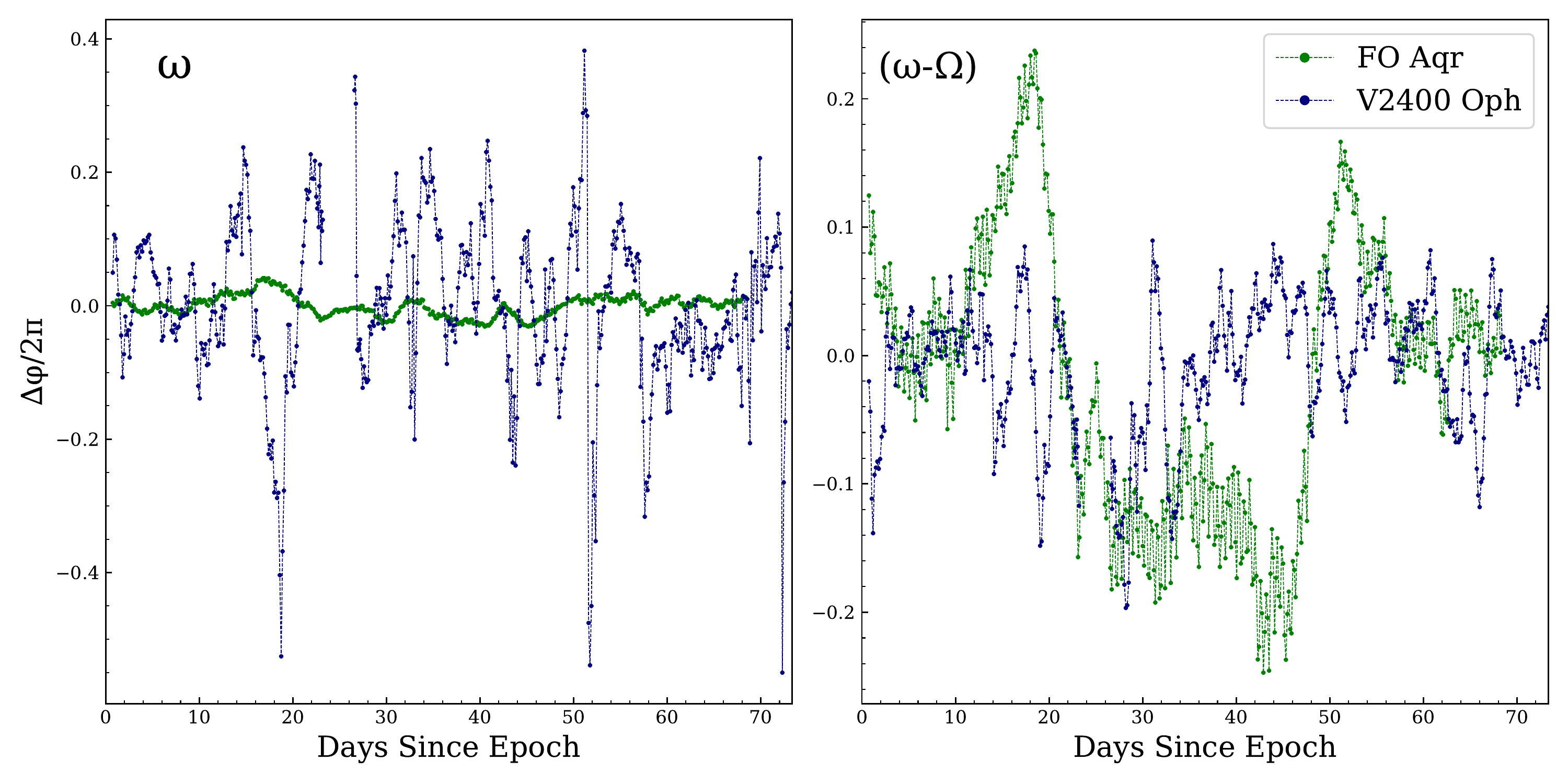}
    \caption{Relative phase change for spin and beat signals over the duration V2400 Oph and FO Aqr K2 observations. Each data point corresponds to a 1.5-d observation window. The FO Aqr spin signal phase is by far the most stable showing minimal variation compared to the other three signals. The FO Aqr and V2400 Oph beat signals show similar levels of phase variability over their respective observations.}
    \label{fig:phase_change}
\end{figure}

A change in the phase of a signal is often associated with a change in the accretion rate. Figure \ref{fig:phase_change} further illustrates the dynamic variability observed in V2400 Oph compared to FO Aqr. The variation in spin phase is relatively stable for FO Aqr. While its beat phase variation is large, the evolution is stable and correlated with the brightness of the system. However, the spin and beat signals in V2400 Oph demonstrate drastic variations over short time scales that are uncorrelated with any other phenomena. This indicates that the spin and beat signals of V2400 Oph have a different physical origin than those of FO~Aqr during their respective \textit{K2} observations.

\subsubsection{TX~Col}
TX~Col was observed by \textit{TESS} during two separate, eight-week spans, a similar observation length as the V2400 Oph K2 observations. During both observations, TX~Col entered a bright state during which the well-behaved, periodic variability in its light curve was abruptly overrun by large-amplitude, low-coherence QPOs, albeit at higher frequencies than observed in V2400~Oph \citep{littlefield2021}. 

In both TX~Col and V2400~Oph, the QPO amplitude has been obvserved to exceed that of periodic variability during epochs of enhanced accretion. Based on the typical ASAS-SN $g$ magnitudes near maximum light for TX~Col and V2400~Oph, their Gaia EDR3 distances \citep{edr3, BJ21}, and estimates of their extinction,\footnote{The \citet{green} three-dimensional dust maps indicate a total extinction of $A_g = 0.81^{+0.03}_{-0.18}$~mag for V2400~Oph, but these maps do not cover TX~Col's position. However, the \citet{sf11} maps indicate the total Galactic extinction along TX~Col's line of sight is only $A_g = 0.13$~mag.} their absolute magnitudes in the $g$ band are 4.9 and 4.3, respectively. These absolute $g$ magnitudes, which are comparable to the absolute $V$ magnitudes of non-magnetic dwarf novae during outburst \citep{warner87, patterson2011}, suggest that both TX~Col and V2400~Oph have high accretion rates. Likewise, during episodes of reduced accretion, the light curves of TX~Col and V2400~Oph have shown well-defined pulses at $(\omega-\Omega)$ \citep{Buckley1995Discovery, littlefield2021}, a sharp contrast with the erratic light curves that characterize their bright states. 

The comparison with TX~Col is not perfect. Most prominently, the amplitude of the TX~Col QPOs is largest between 10-25~cycles~ d$^{-1}$, and although these QPOs disappear during epochs of reduced mass transfer, they return at elevated mass-transfer rates, a trend that has existed for several decades \citep{littlefield2021}. In contrast, the QPOs in V2400~Oph (Figure~\ref{fig:ls_periodogram}) occur at much lower frequencies ($\sim2-9$~cycles~d$^{-1}$), and it is unclear whether their presence is as dependent on the mass-transfer rate as their counterparts in TX~Col.

\begin{figure}[t]
    \centering
    \includegraphics[width=\linewidth]{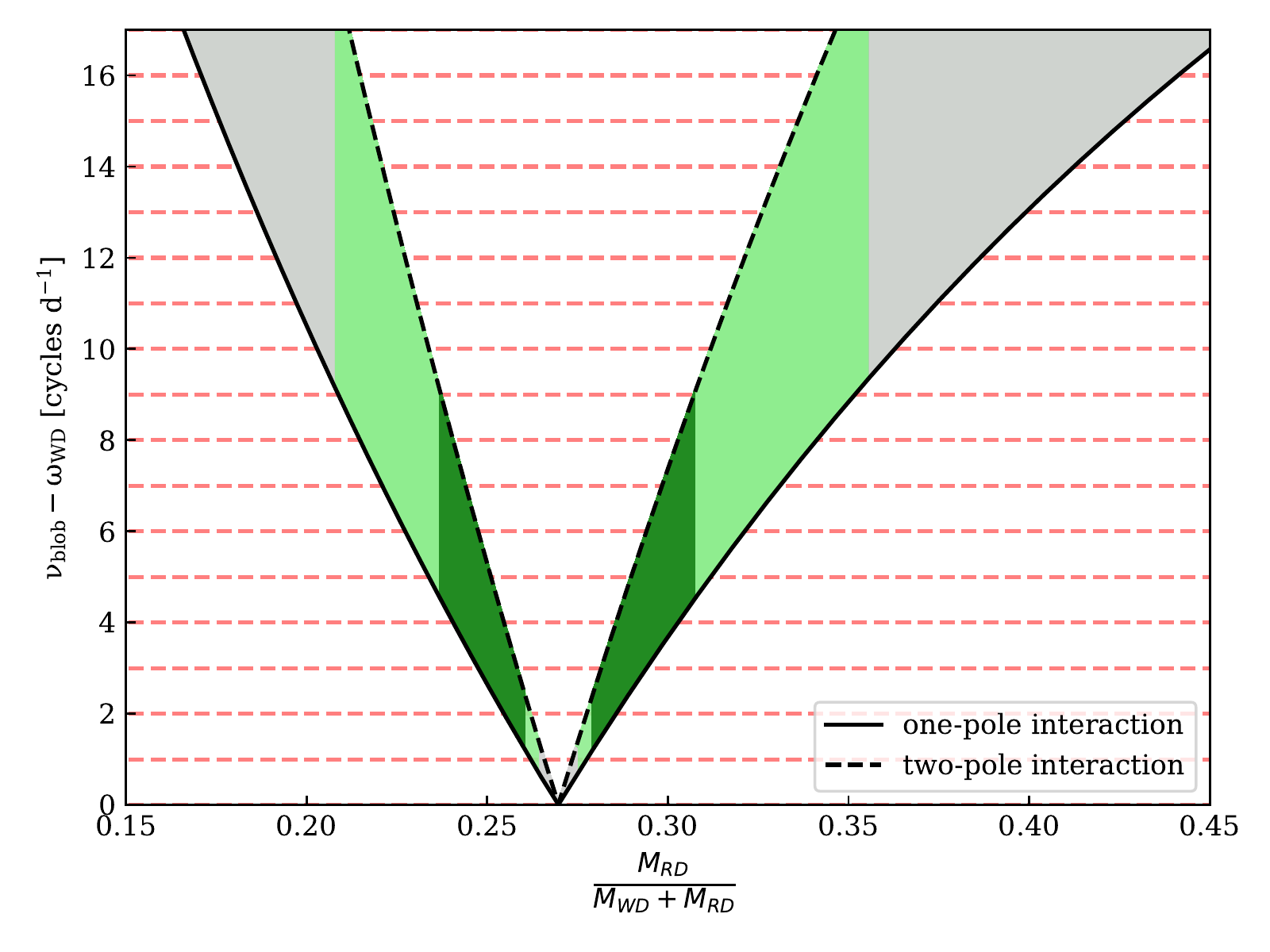}
    \caption{The shaded regions in the figure above represent the predicted beat frequencies between the WD spin frequency and diamagnetic blobs orbiting at the circularization radius with a Keplerian orbit. Because the masses of the stellar components are unknown, we parameterized the predicted frequency against a range of plausible stellar mass ratios. The two limiting cases, blobs interacting with one pole or interacting with both poles equally, defines the range of plausible frequencies at each mass ratio. The dark green region represents the upper and lower bound of V2400 Oph's QPOs (2-9~cycles~d$^{-1}$) produced through two-pole interaction while the light green region defines the range for one-pole interaction. 
    }
    \label{fig:QPO_model}
\end{figure}

\citet{littlefield2021} proposed the TX~Col QPOs were caused by individual diamagnetic blobs near the binary's circularization radius beating against the WD spin frequency. Their calculations showed this phenomenon could produce the observed TESS TX~Col QPO frequencies for plausible values of the stellar masses. The model presented by \cite{littlefield2021} calculates the difference between the WD spin frequency and Keplerian frequency at the circularization radius, requires four input parameters: $P_{\omega}$, $P_{\Omega}$, and the stellar masses M$_{1}$ and M$_{2}$.  $P_{\omega}$ and $P_{\Omega}$ are well-known for V2400~Oph, but M$_{1}$ and M$_{2}$ are not. To determine the viability of a similar effect in V2400 Oph, we calculated the Keplerian frequency at the circularization radius for a range of plausible stellar mass ratios. The beat frequency between each Keplerian frequency and the WD spin frequency is shown in Figure~\ref{fig:QPO_model}. We find that for typical donor-star masses at $P_{\Omega}=3.4$~h ($0.2-0.3~\mathrm{M}_{\odot}$) and an adopted WD mass of 0.8~M$_{\odot}$, the predicted QPO frequencies are quite low ($\lesssim10$~cycles~d$^{-1}$). In these solutions, the blob orbital frequency would be below the local field and it shows that for plausible stellar masses, the expected beat frequencies between the WD spin and blob orbital frequencies are quite low, consistent with the \textit{Kepler} observations.

\subsection{Testing Theoretical Power Spectral Modeling}

Noting V2400 Oph as the best candidate for disk-less accretion, \citet{FerrarioModel} presented theoretical power spectra for stream-fed and disk-fed IPs using a model that accounts for optical variations in the accretion curtain and funnel. While \citet{FerrarioModel} provided no optical continuum power spectra for V2400 Oph, we can infer predictions based their the generalized optical power spectra for a stream-fed accretor. The power spectrum in Figure~\ref{fig:ls_periodogram} best agrees with a theoretical power spectrum that has low inclination and high field asymmetry parameters. \citet{FerrarioModel} discussed that for stream-fed accretors, a high asymmetry parameter indicates pole flipping is present in the system. This interpretation implies the presence of V2400 Oph possessing pole-flipping as discussed previously by \citet{Hellier2002} using X-ray observations. The \citet{FerrarioModel} asymmetry parameter allows for the two poles to contribute unequally to the light curve, as might be expected with pole flipping so it would provide a possible explanation for only 25\% of accretion material participating in pole flipping as reported in \cite{Hellier2002}.

Although stream-fed modeling has success in describing the beat signal observed, the dominance of low-frequency QPOs in V2400~Oph is not predicted by the \citet{FerrarioModel} modeling. It is only when observed for long period durations that V2400 Oph reveals its periodic nature. The simulated strem-fed continuum flux light curve in \citet{FerrarioModel} is in no way representative of the features seen in Figure \ref{fig:lightcurve_segments}. Since the \cite{FerrarioModel} modeling only accounts for contributions from accretion streams, it is plausible that V2400 Oph possesses other accretion mechanisms behind its aperiodic and QPO dominated characteristics.

\subsection{The Case for Diamagnetic Blob Accretion}
 The K2 light curve of V2400~Oph is dominated by aperiodic behavior and low-frequency quasi-periodic oscillations. Where standard IP accretion theories fail to predict this behavior, the diamagnetic blob model (DBM) may offer solutions. The DBM was first introduced by \cite{King1993} and later numerically expanded in \citet{WynnKing1995}. Past literature on V2400~Oph has invoked the DBM to explain the system's peculiar properties \citep{Hellier2002, Martino2004, Hellier2014, Joshi2019}. It should be noted that \citet{Norton2004, Norton2008}, have since expanded the range of accretion models numerically into four categories (streams, rings, disks, propellers) and suggest mechanisms may coexist in a system. The DBM proposed by \citet{King1993} seems to capture that coexistence. From the analytic treatment of the DBM in \citet{King1993}, three points of accretion physics emerge involving $E_{blob}$, which is defined by a discrete blob's mechanical energy per unit mass, $E_{blob}= \frac{1}{2}v^2 - \frac{GM_{wd}}{r}$.

\begin{enumerate}
    \item Only blobs below a critical energy, $E_{acc}$, will undergo accretion onto the WD. Blobs with $E_{blob} > E_{acc}$ will gain angular momentum through interacting with the WD's spinning magnetosphere and re-accrete onto the companion star.
    
    \item Blobs with $E_{blob} < E_{acc}$ will circularize near the circularization radius, $R_{circ}$, due to magnetic drag. The WD field strength and blob density will determine if the blobs are threaded by the magnetosphere or self interact to form a disk, $t_{mag}\ \text{vs}\  t_{visc}$. 
    
    \item For cases with field asymmetry, resonances between blob orbits and the magnetic drag may cause the accretion flow to deviate from the orbital plane near the corotation radius, inducing oscillations at $(\omega - \Omega)$.
\end{enumerate}

In light of the K2 observations and using these three propositions, the bulk of the phenomena observed in V2400~Oph can be addressed. The fate of a blob in V2400 Oph is destined for three possible outcomes based on its energy and density. \cite{King1993} predicts blobs for which $E_{blob} > E_{acc}$, angular momentum will be added as the blob interacts with the magnetosphere in an outward spiraling orbit. The low-frequency QPOs may be linked to the optical fluctuations caused by blobs with $E_{blob} > E_{acc}$ beating against the WD spin frequency as move outward along spiraling orbits. The dominant QPOs occur preferentially between 2-9 cycles~ d$^{-1}$ indicating a frequency-varying oscillator, such as blobs moving along and expelled from the outer regions of the Roche lobe. Given the predominant power at these frequencies, we expect that most of the  blobs in V2400~Oph follow this trajectory. 

Outcomes for accreting blobs with $E_{blob} < E_{acc}$ are determined by their densities and resulting $t_{mag}$. Our presumption is that the strong magnetic field in V2400~Oph creates a situation where  $t_{mag} \sim t_{visc}$ for the accreting blobs. The evolving distribution of blob densities results in variable accretion mode dominance as observed in Figure \ref{fig:signal_modulation}. As denser, longer-orbiting blobs enter the magnetosphere their $t_{mag} > t_{visc}$, so they circularize around $R_{circ}$ and briefly self-interact through viscous forces and may even cause disk formation. The interpretation of Figure \ref{fig:loglog_periodogram}, provides support for blobs interacting through viscous forces at the observed `flickering frequencies'. Further, an incoherent spin signal indicates the number density of circularized blobs is variable, as expected for an unpredictable process like diamagnetic blob accretion. 

In the last scenario, low-density blobs, $t_{mag} << t_{visc}$, will be threaded by the magnetosphere and accrete in a stream-like manner. As a result, pole-flipping causes optical fluctuations at  $(\omega - \Omega)$. Based on theoretical power spectra in \citet{FerrarioModel}, V2400 Oph demonstrates strong evidence for stream-fed accretion in an asymmetric field. As noted above, \citet{FerrarioModel} predicts that field asymmetries will produce oscillations at the beat frequency. As the number density of $t_{mag} < t_{visc}$ blobs present in the system evolves, the power at $(\omega - \Omega)$ will be intermittent -- a feature demonstrated in Figure \ref{fig:trailed_power} and Figure \ref{fig:signal_modulation}.

Based on the relative power of $(\omega - \Omega)$, $(\omega)$, and QPOs we can say something about the kind of blobs coming from the companion star. We expect that most blobs have $E_{blob} > E_{acc}$ and spiral out creating QPOs. Of the accreting blobs, we find that low density, stream-accreting blobs must be predominant, yet accrete in bulk and at a rate which is faster than they can be consistently replenished. This may explain why the beat signal amplitude is highly variable over short time-scales as seen in Figure \ref{fig:signal_modulation}. In comparison, the dense viscous blobs are more stable resulting in smaller variations in spin signal over the observation. When low density blobs are absent from the system, the longer lasting dense blobs will allow the spin to prevail as the dominant signal. As mentioned in Sec \ref{Sec: Geometry}, the spin amplitude may be damped due to the observation angle and may otherwise be comparable to the beat amplitude. This would change the expected relative number of high density accreting blobs within the system. Overall, the combination of these three blob scenarios produces a messy system as captured by the aperiodic fluctuations distinctly observed in the light curves of V2400~Oph. 

We encourage further theoretical modeling of V2400 Oph in light of the K2 observations and proposed diamagnetic blob accretion.

\acknowledgments
We thank the anonymous referee for their timely report and highlighting later contributions to the DMB model. AL recognizes the support of the Flatley~CUSE~IGNITE~Fellowship. PS acknowledges support from NSF grant AST-1514737. M.R.K. acknowledges support from the ERC under the European Union’s Horizon 2020 research and innovation programme (grant agreement No. 715051;Spiders).

\facility{Kepler}
\software{Matplotlib \citep{matplotlib},
NumPy \citep{numpy}, Astropy \citep{astropy:2013, astropy:2018}, Lightkurve \citep{lightkurve}}

\appendix 

In the course of reanalyzing the FO Aqr data, we created a two-dimensional power spectrum with a significantly narrower window than the one used in \citet{Kennedy2016}. Our 2D power spectrum does not alter any of the conclusions by \citet{Kennedy2016}, but we present it here because it offers additional insight into the evolution of FO Aqr's power spectrum during its \textit{K2} observation. Furthermore, it offers a strong comparison with our 2D power spectrum of V2400 Oph in Figure \ref{fig:trailed_power}.

\begin{figure*}[h]
    \centering
    \includegraphics[width =0.92\textwidth]{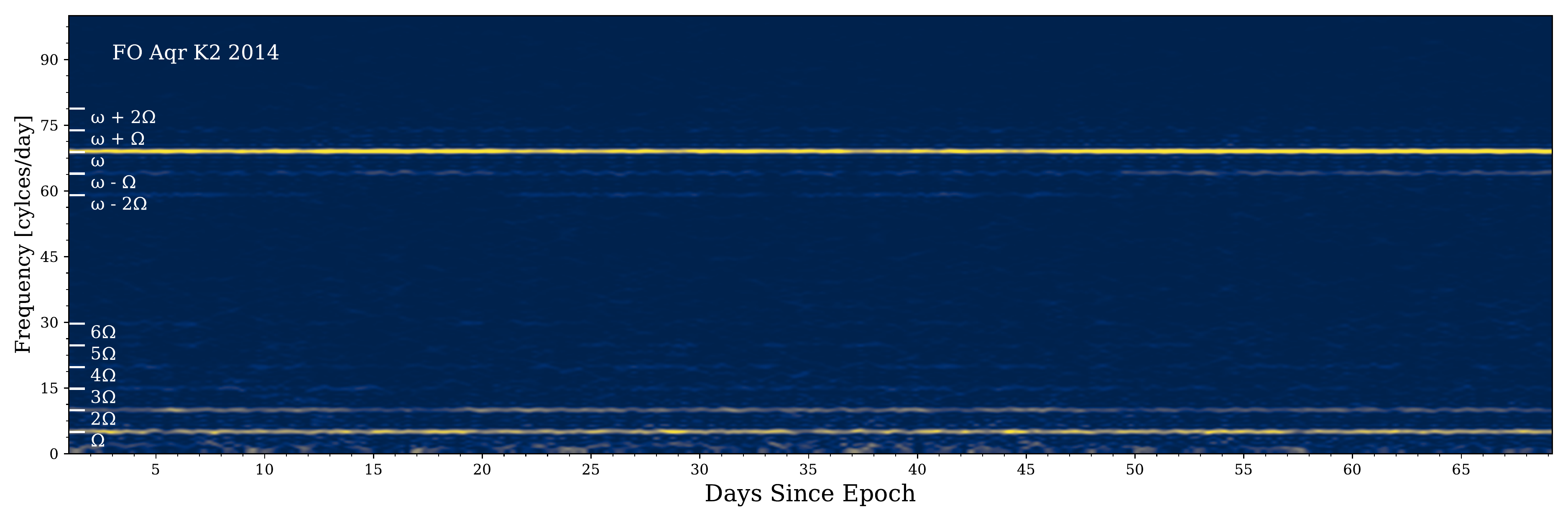}
    \caption{FO Aqr's K2 power spectrum binned at 1.25 days with .25 day increments. The spin and orbital frequencies are seen at the highest powers throughout the observation. $( T_0 \approx 2144.0 \ \text{BKJD})$}
   \label{fig:fo_trailpower}
\end{figure*}

\bibliography{references.bib}

\end{document}